# Coordinated Expansion Planning of Transmission and Distribution Systems Integrated with Smart Grid Technologies

Mojtaba Moradi-Sepahvand[*], Turaj Amraee, Farrokh Aminifar, and Amirhossein Akbari

*Abstract*— Integration of smart grid technologies in distribution systems, particularly behind-the-meter initiatives, has a direct impact on transmission network planning. This paper develops a coordinated expansion planning of transmission and active distribution systems via a stochastic multistage mathematical programming model. In the transmission level, in addition to lines, sitting and sizing of utility-scale battery energy storage systems and wind power plants under renewable portfolio standard policy are planned. Switchable feeders and distributed generations are decision variables in the distribution level while the impact of demand response programs as a sort of behind-the-meter technologies is accommodated as well. Expansion of electric vehicle taxi charging stations is included as a feasible option in both transmission and distribution levels. In order to deal with short-term uncertainty of load demand, renewable energy sources output power, and the charging pattern of electric vehicle taxis in each station, a chronological time-period clustering algorithm along with Monte Carlo simulation is utilized. The proposed model is tackled by means of Benders Dual Decomposition (BDD) method. The IEEE RTS test system (as the transmission system) along with four IEEE 33-node test feeders (as distribution test systems) are examined to validate effectiveness of the proposed model.

__________

Mojtaba Moradi-Sepahvand is with the Department of Electrical Sustainable Energy, Delft University of Technology, Delft, The Netherlands (e-mail: m.moradisepahvand@tudelft.nl).
Turaj Amraee, and Amirhossein Akbari are with the Faculty of Electrical Engineering, K.N. Toosi University of Technology, Tehran, Iran (e-mails: amraee@kntu.ac.ir, and a.h.akbari@email.kntu.ac.ir).
F. Aminifar is with the School of Electrical and Computer Engineering, College of Engineering, University of Tehran, Tehran, Iran (e-mail: faminifar@ut.ac.ir).
*Corresponding author.
Mojtaba Moradi-Sepahvand, E-mail: m.moradisepahvand@tudelft.nl

# NOMENCLATURE

*Acronyms:*

| | |
|---|---|
| BDD, BES | Benders Dual Decomposition, Battery energy storage. |
| CRF, CTPC | Capital recovery factor, Chronological time-period clustering. |
| DDSP, DEP, DG | Distribution dual sub-problem, Distribution expansion planning, Distributed generation. |
| DPV, DRPs, DSO | Discounted present values, Demand response programs, Distribution system operator. |
| EAC, EVs, EVCSs | Equivalent annual costs, Electric vehicles, Electric vehicle charging stations. |
| EVTs, EVTCSs | Electric vehicle taxis, Electric vehicle taxis charging stations. |
| LMPs, MCS, MP | Local marginal prices, Monte Carlo simulation, Master problem. |
| RESs, RPS | Renewable portfolio standard, Renewable energy sources. |
| TDIC, TDOC, TDGs | Total distribution investment costs, Total distribution operation costs, Thermal DGs. |
| TDSP, TEP, TSO | Transmission dual sub-problem, Transmission expansion planning, Transmission system operator. |
| TTIC, TTOC | Total transmission investment cost, Total transmission operation cost. |
| WDGs, WPPs | Wind power DGs, Wind power plants. |

*Indices & Sets:*

| | |
|---|---|
| $c, \Omega_C, m, \Omega_M$ | Index and set of the allowable candidate lines in a transmission corridor, index and set of in-service EVT chargers. |
| $d, \Omega_D$ | Index and set of distribution systems. |
| $f, \Omega_F, \Omega_{nf}, \Omega_{ef}$ | Index and set of all distribution feeders, and sets of all candidate and existing feeders. |
| $h, hd, \Omega_H, H$ | Indices, set, and total number of representative hours. |
| $i, \Omega_B, \Omega_G$ | Index and set of buses, and set of generators in transmission. |
| $l, L$ | Index of all transmission lines, and total number of lines. |
| $n, \Omega_N, N, \Omega_{NN}, nd, \Omega_{ND}$ | Index and set of all nodes, total number of existing nodes, set of new nodes in each distribution system, and index and set of responsive load nodes. |
| $t, \Omega_T, T$ | Index, set, and total number of planning stages. |
| $tp, \Omega_{TP}$ | Index and set of all types of BES devices in transmission system. |
| $s, \Omega_S, S$ | Index, set, and total number of linear segments for generation cost function linearization. |
| $zd, \Omega_{ZD}, \Omega_{ns}$ | Index and set of EVTCSs candidate service zones, and set of candidates EVTCSs in each zone of distribution. |
| $ze, \Omega_{ZE}, \Omega_{cs}$ | Index and set of EVTCSs candidate service zones, and set of candidates EVTCSs in each zone of transmission. |
| $\Omega_{nl}, \Omega_{nc}, \Omega_{el}$ | Sets of all candidate lines, candidate lines in new corridors, and existing lines in transmission. |
| $\Omega_{nw}, \Omega_{nt}$ | Sets of nodes for wind and thermal distributed generation installation in distribution. |
| $\Omega_{sb}, \Omega_w$ | Sets of candidate buses for installing BES and constructing WPPs in transmission. |

*Parameters:*

| | |
|---|---|
| $A, K, IN$ | Connectivity matrices of existing and new lines (or feeders) with buses (or nodes), and discriminant matrix of distribution systems interface nodes with transmission. |
| $Cg_i^s, Cdg_n^s, Cwc_i$ | Segmental generation cost for thermal units and DGs and penalty cost of wind curtailment ($/MWh). |
| $Cs_{i,tp}, Cc_{i,tp}$ | BES investment cost of energy capacity ($/MWh) and power capacity ($/MW). |
| $C_{i,tp}, S_{i,tp}$ | Maximum power (MW) and energy (MWh) capacity of BES. |
| $CR_m, EZ_{i,ze,h}^{TS}, EZ_{d,n,zd,h}^{DS}$ | EVTs charging rate (MW), and discriminant matrices of operational hours of EVTCSs in candidate service zones of transmission and distribution systems. |
| $ES_{i/(n),ze/(zd)}, EC_{i/(n),ze/(zd)}, EP_{i/(n),ze/(zd)}$ | Investment costs of EVTCSs ($), each charger ($), and maximum number of chargers for each station $i$ (or $n$), in zone $ze$ (or $zd$) of transmission (or distribution). |
| $IW_i, DW_n, DT_n$ | Investment cost of new WPP in transmission (M$/MW), and new wind (M$/MW) and thermal (M$) DGs in distribution. |
| $IC_l, IF_f$ | Line $l$ investment cost including the cost of conductors and towers, and new feeder investment cost (M$/km). |
| $LL_l, FL_f$ | New transmission lines and distribution feeders length (km). |
| $LT, r, TL$ | Equipment lifetime (year), interest rate, and upper limit for new TDGs in each distribution system and in each stage. |
| $Ld_i^{PK}, Ld_{d,n}^{PK}, QLd_{d,n}^{PK}, Lg_t, Eg_t$ | Active peak load of buses, active and reactive peak load of nodes (MW), load demand and EVT growth factors. |
| $\mathcal{M}, \Psi, B, G, SF_f$ | Big-M, system base power (MVA), susceptance and conductance of lines and feeders in per unit, and maximum apparent power flow of feeders (MVA). |
| $RU_i, RD_i$ | Ramp up and ramp down of thermal units (MW/hr). |
| $Rw_l, DR_f$ | Cost of right of way for transmission line $l$ consists of land cost, and distribution feeder $f$ (M$/km). |
| $SoC^{DT}, SoC^{AT}$ | EVTs state of charge at departure and arrival hours. |
| $TSb_l, DSb_n$ | Substation cost in new transmission corridors, and in new distribution nodes (M$). |
| $Wf_h, Lf_h, Ef_h$ | Hourly factors obtained for WPP output power, load demand, and EVTs charging pattern. |
| $\alpha, \beta, \chi$ | Expected share of WPP (or WDG) in total load supplying at the end of planning horizon, maximum wind curtailment in each stage, and flexible ramp reserve cost factor. |
| $\gamma_h^{hd}, \mathcal{L}_{d,t,h}, A\mathcal{L}_{d,t}$ | Cross-hour price elasticities, hourly LMPs in interface nodes, and average of hourly LMPs in interface nodes. |
| $\eta_c, \eta_d$ | Charging and discharging efficiency of BES devices. |
| $\rho_h, TE_h$ | The weight of obtained representative hour $h$, and hourly average travel energy of EVTs (MWh). |
| $\{\bullet\}^{max}, \{\bullet\}^{min}$ | Maximum/minimum limits of bounded variables. |

*Variables:*

| | |
|---|---|
| $D_{d,t,nd,h}$ | DRP participation value (MW) in responsive load nodes $nd$, in stage $t$, and hour $h$ in distribution $d$. |
| $DU_{d,t,nd,h}, DL_{d,t,nd,h}$ | Upper and lower bound of DRP (MW) in responsive load nodes $nd$, in stage $t$, and hour $h$ in distribution $d$. |
| $E_{t,i,tp,h}$ | Energy level (MWh) of BES with type $tp$, in bus $i$, in stage $t$, and hour $h$. |
| $I_{t,i,tp}, U_{t,i,tp,h}$ | Binary variables of new BES with type $tp$, in bus $i$, in stage $t$, and the state of charging or discharging of BES with type $tp$, in bus $i$, in stage $t$, and hour $h$. |
| $J_{t,i,h}, XG_{d,t,n}$ | Thermal generation units on/off state, and binary variable of candidate TDG $n$, in stage $t$, and distribution system $d$. |
| $P_{t,i,h}, PS^{TS}_{t,i,h,s}, PTD_{d,t,h}$ | Total and segmental power of thermal units (MW), and power exchange between transmission and distribution system $d$ in interface nodes (MW). |
| $Pd_{t,i,tp,h}, Pc_{t,i,tp,h}$ | Power of discharging and charging of BES with type $tp$, in bus $i$, in stage $t$, and hour $h$ (MW). |
| $Pe^{TS}_{t,l,h}, Pe^{DS}_{d,t,f,h}$ | Active power flow of existing lines and feeders (MW). |
| $PG_{d,t,n,h}, QG_{d,t,n,h}, PS^{DS}_{d,t,n,h,s}$ | Total active (MW) and reactive (Mvar) power, and segmental active (MW) power generation of TDGs. |
| $Pl_{t,l,c,h}, Pf_{d,t,f,h}$ | Active power flow of new lines and feeders (MW). |
| $Pw_{t,i}, PC_{t,i,h}, Pwg_{d,t,n}$ | Power capacity and wind curtailment (MW) of WPPs, and power capacity of WDG (MW). |
| $Qe_{d,t,f,h}, Qf_{d,t,f,h}$ | Reactive power flow of existing and new feeders (Mvar). |
| $R_{t,i,h}$ | Reserve of thermal unit $i$, in stage $t$, and hour $h$ (MW). |
| $XE^{TS}_{t,i,ze,h,m}, XE^{DS}_{d,t,n,zd,h,m}$ | Binary variables of in-service EVT charger $m$, in EVTCS $i$ of zone $ze$, in stage $t$, and hour $h$ of transmission, and EVT charger $m$, in EVTCS $n$ of zone $zd$, in stage $t$, and hour $h$ in distribution $d$. |
| $XO_{d,t,f,h}, XN_{d,t,n}$ | Binary variables of open/close state of feeder $f$, in stage $t$, and hour $h$ in distribution $d$, and new node $n$, in stage $t$ in distribution $d$. |
| $XS^{TS}_{t,i,ze}, XS^{DS}_{d,t,n,zd}$ | Binary variables of candidate EVTCS $i$, in stage $t$, and zone $ze$ of transmission, and candidate EVTCS $n$, in stage $t$, and zone $zd$ in distribution $d$. |
| $Y_{t,l,c}, X_{d,t,f}$ | Binary variables of candidate line $l$, in stage $t$, and corridor $c$, and candidate feeder $f$, in stage $t$ in distribution $d$. |
| $Z$ | Total Planning Cost. |
| $ZD^+_{d,t,nd,h}, ZD^-_{d,t,nd,h}$ | Binary variables of upper and lower bound of DRP in responsive load nodes $nd$, in stage $t$, and hour $h$ in distribution $d$. |

| $\theta_{t,i,h}, V_{d,t,n,h}$ | Voltage angle of bus *i*, in stage *t*, and hour *h*, Voltage magnitude of node *n*, in stage *t*, and hour *h* in distribution *d*. |

*Lump Symbols:*

| $P^{TS/DS}$ | Positive continuous variables vector. |
| $Q^{TS/DS}$ | Free continuous variables vector. |
| $W^{TS/DS}$ | WPP and WDG power capacity variables vector. |
| $Y^{TS/DS}$ | Binary decision variables vector. |

## I. INTRODUCTION

In conventional passive distribution systems, the main focus is on distributing a predetermined amount of power from transmission substations to the medium and low voltage load centers. Accordingly, the operation and planning of transmission and distribution systems have been regularly conducted separately [1]. In the wake of distribution system reformation to host high penetration of distributed generation (DG), demand response programs (DRPs), and high shares of electric vehicles (EVs), the local supply of load demands is turning into a reality in the active distribution networks. This fundamental transition enables the distribution system to run as dispatchable sources and dramatically influences the net load demand profiles. In a rare but still feasible scenario, the extra power generated in distribution side can be injected to the main grid under an interactivity between transmission system operator (TSO) and distribution system operator (DSO). This bilateral power exchange, in addition to power and energy quantities, impacts the prices and monetary flows, and the decisions in transmission expansion planning (TEP) should be amended accordingly. Also, high penetration of wind power plants (WPPs) in transmission system and integration of flexible resources like utility-scale battery energy storage (BES) systems, can affect distribution expansion planning (DEP). In this regard, a proper coordination between transmission and distribution systems planning is essential to defer investment and rise the asset utilization [1].

In traditional expansion planning models, the focus is mainly on a particular part of power system, e.g., transmission grid, or distribution system. In [2] the co-planning of transmission grid and ES devices is addressed, and the importance of ES for relieving lines congestion is concluded. Authors of [3] and [4] have developed an expansion model under high penetration of renewable energy sources (RESs) considering power system uncertainties. In [5], a hybrid AC/DC TEP model under high penetration of RESs incorporating BES devices is presented. The research works proposed in [6-8] investigate the impact of EVs [6], or DRPs [7, 8] on power system operation and planning. An interlink between EV routing and optimal charging strategy in power system is developed in [6]. The capability of DRPs for cost and emission reduction in generation and transmission expansion is addressed in [7]. A power system expansion framework considering internet data centers load regulation is presented in [8] for facilitating the inclusion of spatial and chronological internet data centers in DRPs. In [9-14] the expansion of distribution systems considering some smart technologies is investigated. A mathematical linearization for a reliability-based DEP problem is presented in [9]. The co-planning of electric vehicle charging stations (EVCSs), ES devices, and DGs, in a radial distribution network is developed in [10]. Authors of [11] propose a model for expansion planning of distribution and transportation systems, considering ES, DGs, and shared EVCSs, for minimizing investment cost, energy losses, and queue waiting time of EVs. In [12], a stochastic model considering EVs charging demand is utilized for active distribution system reinforcement planning. In [13], optimal expansion of distribution system and EVCSs assuming coordinated power and transportation networks is addressed. In [14] the presented model in [13] is expanded by including ES and fast-charging stations.

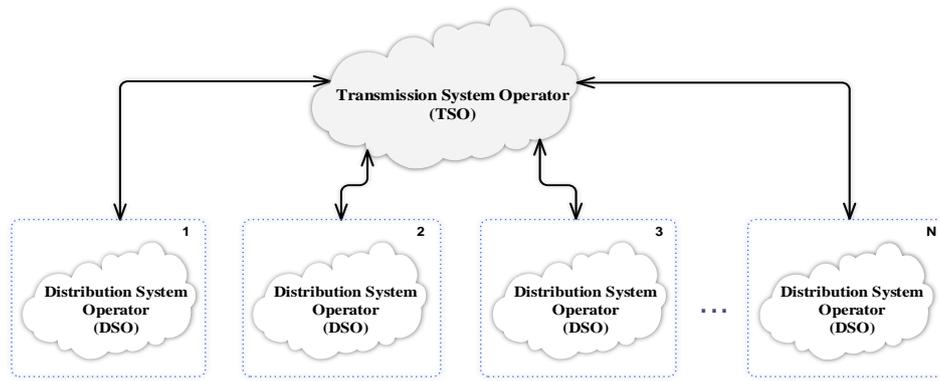

Fig. 1. The bidirectional interaction between TSO and DSO

Followed by growing penetration of RESs, modern expansion tools, and smart grid technologies it is crucial to further look into the bidirectional interaction and coordination between TSO and DSO, as illustrated in Fig. 1. In this regard, a short-term study is conducted on sharing energy storage between TSO and DSO in [15] where a coordinated local ES system in distribution level is modeled to relieve transmission system congestions. A coordinated economic dispatch of transmission and distribution systems has been addressed in [16]. In [17], the impact of optimal operation of distribution system on TEP problem is investigated ignoring the optimal siting and sizing of BES devices and WPPs. Unlike [16] and [17], an integrated model for TEP and DEP is proposed in [1]. The sitting and sizing of utility scale BES devices and WPPs, along with EVCSs in transmission side, also penetration of EVs and applications of DRPs in distribution side are all ignored in [1]. In [18] a bi-level framework for coordinated planning of transmission and distribution systems considering the penetration of DGs is presented. The ES devices, WPPs, EVs, and DRPs are not included in [18]. A tri-level structure is considered in [19] for TSO, DSO, and independent system operator to coordinate TEP and DEP problems. In [20] a hierarchical collaborative structure for TEP and DEP is addressed considering the allocation of transmission cost. The uncertainties, ES, WPPs, EVs, and DRPs are all ignored in [19] and [20]. In [21] an integrated model for planning of power system and fast charging stations under EV diffusion is presented. DRPs impact in distribution systems, and utility scale ES and WPPs, along with EVCS planning in transmission system are not considered in [21].

The EV market in active distribution systems based on the stated policies scenario (STEPS) is targeting 140 million EV penetration by 2030, compared to 7.2 million existing ones in 2020 [22]. This rapid growth, which will dominate 7% of the global vehicle fleet, can cause a 550 TWh increase in global electricity demand in the same scenario and time [22]. Electric vehicle taxis (EVTs) have a growing penetration in distribution systems. Due to different consumption patterns and long charging time, EVTs need the EVTs-specific charging stations [23]. Therefore, EVT charging stations (EVTCSs) introduce a new challenge for the operation and expansion planning of power systems. Although the proposed models in [1, 15-20] have incorporated an interaction between transmission and distribution systems, inclusion of EVTs or EVs can give more flexibility to the problem. Moreover, in [21] just the impact of private EVs in distribution systems is investigated. The impacts of EVs charging and solar generated power with respect to TSO and DSO interaction in a short-term horizon is addressed in [24] ignoring ES devices and WPPs in transmission, and DRP in distribution. Furthermore, DRP is an option in active distribution systems for electric energy consumers to contribute in the power system operation. In [25] DRP impacts on DEP problem is investigated. In Table I, the previous papers are briefly reviewed in terms of the planning model description and considered decision variables in transmission and distribution sides.

As shown in Table I, in majority of previous studies (e.g., [2-5, 7-14, 25]), the TEP and DEP problems are conducted separately. Moreover, in previous coordinated TEP and DEP research works (e.g., [1, 18-21]), the sitting and sizing of utility scale BES

Table I. An overview of the previous papers

| Ref[1] | Model Description | | | Transmission Side Tools | | | | Distribution Side Tools | | | |
|---|---|---|---|---|---|---|---|---|---|---|---|
| | TS[2] | DS[3] | TS&DS | TL[4] | ES[5] | WPP[6] | EVCS[7] | SF[8] | DG[9] | DRP[10] | EVCS |
| [2], [5] | ✓ | --- | --- | ✓ | ✓ | --- | --- | --- | --- | --- | --- |
| [3] | ✓ | --- | --- | ✓ | --- | ✓ | --- | --- | --- | --- | --- |
| [4] | ✓ | --- | --- | ✓ | ✓ | ✓ | --- | --- | --- | --- | --- |
| [7] | ✓ | --- | --- | ✓ | --- | ✓ | --- | --- | --- | ✓ | --- |
| [8] | ✓ | --- | --- | ✓ | --- | --- | --- | --- | --- | ✓ | --- |
| [9] | --- | ✓ | --- | --- | --- | --- | --- | ✓ | --- | --- | --- |
| [10], [11], [12] | --- | ✓ | --- | --- | --- | --- | --- | ✓ | ✓ | --- | ✓ |
| [13], [14] | --- | ✓ | --- | --- | --- | --- | --- | ✓ | --- | --- | ✓ |
| [25] | --- | ✓ | --- | --- | --- | --- | --- | ✓ | ✓ | ✓ | --- |
| [1], [18], [19], [20] | --- | --- | ✓ | ✓ | --- | --- | --- | ✓ | ✓ | --- | --- |
| [21] | --- | --- | ✓ | ✓ | --- | --- | --- | ✓ | ✓ | --- | ✓ |
| **Proposed Model** | --- | --- | ✓ | ✓ | ✓ | ✓ | ✓ | ✓ | ✓ | ✓ | ✓ |

**1**: Reference, **2**: Transmission System, **3**: Distribution System, **4**: Transmission Line, **5**: Energy Storage, **6**: Wind Power Plant, **7**: Electric Vehicle Charging Station, **8**: Switchable Feeder, **9**: Distributed Generation, **10**: Demand Response Program.

devices and WPPs in transmission side and penetration of EVTs and applications of DRPs in distribution side are not considered. In addition, in previous studies (e.g., [10-12, 21, 24]), mainly the impact of private-owned EVs on power system expansion is considered, and the importance of DRPs in [9-21, 24] is ignored. This paper focuses on short-term and long-term impacts of EVTCSs and DRP on the coordinated expansion planning of the transmission and distribution systems. Moreover, in distribution side DGs are planned for dealing with the increasing load and EVTs charging demand which may lead to a highly cost reduction in the transmission system.

Regarding the literature gaps, in what follows, the main contributions of this paper are summarized.

1) A stochastic multistage model is proposed for coordinated expansion planning of transmission and active distribution systems considering operational details. While the investment decisions can be made by public or private sectors in a national or regional levels, the operational management is handled by TSO and DSO. In the transmission side, in addition to lines, optimal planning of BES devices, and WPPs under renewable portfolio standard (RPS) policy are sought. The expansion of switchable feeders, and DGs are considered as planning options in the distribution side. In both sides the expansion of EVTCSs in several candidate service zones is planned as well.

2) In the proposed coordinated model, with regard to TSO local marginal prices (LMPs) in interface buses, the impact of DRPs implemented by DSO is planned and scheduled.

3) The proposed model is formulated as an MILP problem and is reformulated and solved using a customized Benders Dual Decomposition (BDD) method.

Moreover, to handle the uncertainty of load demand, RESs output power, and EVTs charging pattern in each EVTCS, the chronological time-period clustering (CTPC) algorithm along with Monte Carlo simulation (MCS) are utilized.

## II. Problem Formulations

In the following, the formulations of the proposed coordinated planning model are presented in the integrated and BDD forms.

**A. Objective Function**

In Fig. 2 the considered planning options and smart grid technologies in both transmission and distribution sides are illustrated in a sample interface bus. According to (1), the aim of proposed model objective function is minimizing the Discounted Present Values (DPV) of the Total Transmission Investment Cost (TTIC), Total Transmission Operation Cost (TTOC), and the summation of all Total Distribution Investment Costs (TDIC) along with Total Distribution Operation Costs (TDOC). For all investment costs the DPV is assumed at the middle of each planning stage, and for all operation costs it is considered at the end of each stage. By using Capital Recovery Factor (CRF), the DPVs are converted to Equivalent Annual Costs (EAC) [26]. As presented in (1a), the TTIC consists of new transmission lines, BES devices, EVTCSs with related chargers, and WPPs investment costs. Based on (1b), the investment costs of distribution systems new switchable feeders, substations, thermal DGs (TDGs), EVTCSs with related chargers, and wind power DGs (WDGs) are included in the TDIC. The TTOC includes the linearized cost function of thermal generating units considering required flexible ramp reserve cost, and the total wind curtailment costs as expressed by (1c). Finally, in (1d) the TDOC is defined as the linearized cost function of TDGs.

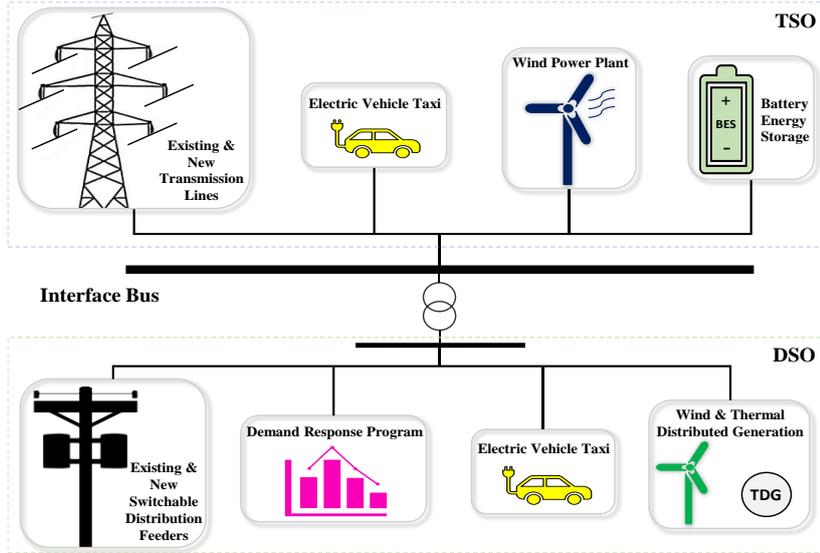

Fig. 2. The planning options and smart grid technologies in the proposed model

$$\mathbf{Min}\ \ \mathbf{Z} = \mathbf{TTIC} + \mathbf{TTOC} + \sum_{d \in \Omega_D}(\mathbf{TDIC}_d + \mathbf{TDOC}_d) \tag{1}$$

$$\mathbf{TTIC} = \sum_{t \in \Omega_T}\ \left[\left(\frac{2}{(1+r)^{2t-1}}\right) \times \right( \tag{1a}$$

$$\frac{10^6 \times r(1+r)^{LT_{Line}}}{(1+r)^{LT_{Line}} - 1} \times \left[\sum_{l \in \Omega_{nl}} \sum_{c \in \Omega_C}(Y_{t,l,c}) \times (Rw_l + IC_l)\cdot LL_l + \sum_{l \in \Omega_{nc}} Y_{t,l,c=1} \cdot (TSb_l)\right] +$$

$$\frac{r(1+r)^{LT_{ES}}}{(1+r)^{LT_{ES}} - 1} \times \sum_{i \in \Omega_{sb}} \sum_{tp \in \Omega_{TP}}(I_{t,i,tp}) \times \left[(Cs_{i,tp}\cdot S_{i,tp}) + (Cc_{i,tp}\cdot C_{i,tp})\right] +$$

$$\frac{r(1+r)^{LT_{EVT}}}{(1+r)^{LT_{EVT}} - 1} \times \sum_{i \in \Omega_{cs}} \sum_{ze \in \Omega_{ZE}}(XS^{TS}_{t,i,ze}) \times \left[ES_{i,ze} + EC_{i,ze}\cdot EP_{i,ze}\right] +$$

$$\frac{10^6 \times r(1+r)^{LT_{WPP}}}{(1+r)^{LT_{WPP}} - 1} \times \left[\sum_{i \in \Omega_W} IW_i\ \cdot Pw_{t,i}\right]\ \Big)\ ]$$

$$TDIC_d = \sum_{t\in\Omega_T} \left[\left(\frac{2}{(1+r)^{2t-1}}\right) \times \right( \quad (1b)$$

$$\frac{10^6 \times r(1+r)^{LT_{fd}}}{(1+r)^{LT_{fd}}-1} \times \left[\sum_{f\in\Omega_{nf}}(X_{d,t,f}) \times (DR_f + IF_f).FL_f + \sum_{n\in\Omega_{nn}} XN_{d,t,n}.(DSb_n)\right] +$$

$$\frac{r(1+r)^{LT_{TDG}}}{(1+r)^{LT_{TDG}}-1} \times \sum_{n\in\Omega_{nt}} DT_n \,.XG_{d,t,n} +$$

$$\frac{r(1+r)^{LT_{EVT}}}{(1+r)^{LT_{EVT}}-1} \times \sum_{n\in\Omega_{ns}} \sum_{zd\in\Omega_{ZD}} (XS^{DS}_{d,t,n,zd}) \times [ES_{n,zd} + EC_{n,zd}.EP_{n,zd}] +$$

$$\frac{10^6 \times r(1+r)^{LT_{WDG}}}{(1+r)^{LT_{WDG}}-1} \times \left[\sum_{n\in\Omega_{nw}} DW_n \,.Pwg_{d,t,n}\right] \left. \right)\right]$$

$$TTOC = \sum_{t\in\Omega_T} \left[\left(\frac{2}{(1+r)^{2t}}\right) \times 8760 \times \sum_{h\in\Omega_H} \rho_h \times \right( \quad (1c)$$

$$\sum_{i\in\Omega_G}\left[\left[Cg_i^{s=1}.(P_i^{min}.J_{t,i,h} + \chi.R_{t,i,h})\right] + \sum_{s\in\Omega_S}[Cg_i^s.PS^{TS}_{t,i,h,s}]\right] + \sum_{i\in\Omega_w} Cwc_i \,.PC_{t,i,h} \left. \right)\right]$$

$$TTOC_d = \sum_{t\in\Omega_T} \left[\left(\frac{2}{(1+r)^{2t}}\right) \times 8760 \times \sum_{h\in\Omega_H} \rho_h \times \left(\sum_{n\in\Omega_{nt}} \sum_{s\in\Omega_S}[Cdg_n^s.PS^{DS}_{d,t,n,h,s}]\right)\right] \quad (1d)$$

## B. Technical Constraints

### B.1. Transmission Level Constraints

In power system expansion planning studies the linear DC optimal power flow (OPF) model is the widely used methodology for modeling power system operational details in a long-term planning model. The DC OPF model calculates the power flows of transmission lines with some simplifications and acceptable accuracy. On the other hand, the AC power flow model has a better accuracy, however; the AC power flow equations are nonlinear. By considering the AC power flow model, the resulted optimization model will be fully non-linear with a large computational burden. In addition, in a non-linear AC power flow the optimality and feasibility of the optimization problem are not guaranteed. Therefore, in this paper the DC OPF model is used for the transmission level constraints.

- *Thermal Generation Units*

The constraints of (2a)-(2d) are defined to model thermal generation units constraints. In (2a), the upper and lower limits of thermal units output power are defined. The thermal units generation cost function is linearized in (2b). According to (2b) and (2c), hourly output power of units is considered as the minimum output power plus the summation of all linear segments of generated power. The constraint (2d) is introduced to model the ramping constraints of all thermal units.

$$P_i^{min} \cdot J_{t,i,h} \leq P_{t,i,h} \leq P_i^{max} \cdot J_{t,i,h} \qquad \forall t \in \Omega_T, i \in \Omega_G, h \in \Omega_H \qquad (2a)$$

$$P_{t,i,h} = P_i^{min} \cdot J_{t,i,h} + \sum_{s \in \Omega_S} PS_{t,i,h,s}^{TS} \qquad \forall t \in \Omega_T, i \in \Omega_G, h \in \Omega_H \qquad (2b)$$

$$0 \leq PS_{t,i,h,s}^{TS} \leq (P_i^{max} - P_i^{min}) \cdot J_{t,i,h}/S \qquad \forall t \in \Omega_T, i \in \Omega_G, h \in \Omega_H, s \in \Omega_S \qquad (2c)$$

$$\begin{cases} P_{t,i,h} + R_{t,i,h} - P_{t,i,h-1} \leq RU_i & \forall t \in \Omega_T, i \in \Omega_G, h \in \Omega_H \\ P_{t,i,h-1} + R_{t,i,h} - P_{t,i,h} \leq RD_i & \forall t \in \Omega_T, i \in \Omega_G, h \in \Omega_H \end{cases} \qquad (2d)$$

- *RPS Policy and Wind Curtailment*

The generation power capacity bounds of installed WPP are introduced in (3a). It is assumed that a minimum of WPP should be available to supply a determined percentage of the total peak load in each stage of planning. According to the RPS policy, the penetration of WPPs in the last stage of the planning horizon is determined as $\alpha\%$ of the total peak load, according to (3b). The constraint (3c) ensures the availability of each installed WPP till the end of planning horizon. The limits of wind energy curtailment in each hour are satisfied using (3d). In (3e), the maximum permitted wind energy curtailment in each stage is defined as $\beta\%$ of the expected output power of available WPPs.

$$0 \leq Pw_{t,i} \leq Pw_i^{max} \qquad \forall t \in \Omega_T, i \in \Omega_w \qquad (3a)$$

$$[\alpha \times t/T] \times (1 + Lg_t)^{2t} \cdot \sum_{i \in \Omega_B} Ld_i^{pk} \leq \sum_{i \in \Omega_w} Pw_{t,i} \qquad \forall t \in \Omega_T \qquad (3b)$$

$$Pw_{t-1,i} \leq Pw_{t,i} \qquad \forall t \in \Omega_T, i \in \Omega_w \qquad (3c)$$

$$0 \leq PC_{t,i,h} \leq Wf_h \cdot Pw_{t,i} \qquad \forall t \in \Omega_T, i \in \Omega_w, h \in \Omega_H \qquad (3d)$$

$$\sum_{i \in \Omega_w} \sum_{h \in \Omega_H} PC_{t,i,h} \leq \beta \times \sum_{i \in \Omega_w} \sum_{h \in \Omega_H} Wf_h \cdot Pw_{t,i} \qquad \forall t \in \Omega_T \qquad (3e)$$

- *Flexible Ramp Reserve*

Flexible ramp reserve is considered to cover the probable forecast errors and handle the uncertainty of load demand and WPP output power according to (4a)-(4c). The limits of thermal units flexible ramp reserve are bounded by (4a) and (4b). In (4c), the lower bound of total hourly reserve is defined as 3% and 5% of the expected system peak load and WPPs output power, respectively [5].

$$0 \leq R_{t,i,h} \leq P_{t,i,h} \qquad \forall t \in \Omega_T, i \in \Omega_G, h \in \Omega_H \quad (4a)$$

$$R_{t,i,h} + P_{t,i,h} \leq P_i^{max} \qquad \forall t \in \Omega_T, i \in \Omega_G, h \in \Omega_H \quad (4b)$$

$$\sum_{i \in \Omega_G} R_{t,i,h} \geq (3\%) \times (1 + Lg_t)^{2t}.Lf_h.\sum_{i \in \Omega_B} Ld_i^{pk} + (5\%) \times \sum_{i \in \Omega_w} Wf_h.Pw_{t,i}$$
$$\forall t \in \Omega_T, h \in \Omega_H \quad (4c)$$

- *Battery Energy Storage*

Generally, BES devices can be charged during low net-load hours and be discharged during heavy net-load conditions. So, these devices can relieve the transmission congestion and postpone or eliminate the generation and transmission planning investment decisions. Under high penetration of WPPs, to postpone new transmission lines construction, relieve existing transmission lines congestion, and minimize wind curtailment, it is essential to incorporate BES devices. In this regard, to model the sitting and sizing of BES devices in the proposed model the constraints of (5a)-(5g) are introduced [4]. The constraints of (5a) and (5b) determine the limits of BES charging and discharging power. In (5c) and (5d), the state of BES charging and discharging is presented. In (5e), the stored energy level in BES devices is defined as the summation of stored energy at the previous hour and the energy exchange at the current hour. The lower and upper bounds of BES stored energy level are considered in (5f). The constraint (5g) ensures the accessibility of each installed BES till the end of planning horizon.

$$0 \leq \eta_c.Pc_{t,i,tp,h} \leq C_{i,tp}.I_{t,i,tp} \qquad \forall t \in \Omega_T, i \in \Omega_{sb}, tp \in \Omega_{TP}, h \in \Omega_H \quad (5a)$$

$$0 \leq {1}/{\eta_d}.Pd_{t,i,tp,h} \leq C_{i,tp}.I_{t,i,tp} \qquad \forall t \in \Omega_T, i \in \Omega_{sb}, tp \in \Omega_{TP}, h \in \Omega_H \quad (5b)$$

$$\eta_c.Pc_{t,i,tp,h} \leq C_{i,tp}.U_{t,i,tp,h} \qquad \forall t \in \Omega_T, i \in \Omega_{sb}, tp \in \Omega_{TP}, h \in \Omega_H \quad (5c)$$

$$1/\eta_d . Pd_{t,i,tp,h} \leq C_{i,tp}.(1 - U_{t,i,tp,h}) \qquad \forall t \in \Omega_T, i \in \Omega_{sb}, tp \in \Omega_{TP}, h \in \Omega_H \quad (5d)$$

$$E_{t,i,tp,h} = E_{t,i,tp,h-1} + (\eta_c . Pc_{t,i,tp,h}) - (1/\eta_d . Pd_{t,i,tp,h}) \qquad \forall t \in \Omega_T, i \in \Omega_{sb}, tp \in \Omega_{TP}, h \in \Omega_H \quad (5e)$$

$$0 \leq E_{t,i,tp,h} \leq S_{i,tp} . I_{t,i,tp} \qquad \forall t \in \Omega_T, i \in \Omega_{sb}, tp \in \Omega_{TP}, h \in \Omega_H \quad (5f)$$

$$I_{t-1,i,tp} \leq I_{t,i,tp} \qquad \forall t \in \Omega_T, i \in \Omega_{sb}, tp \in \Omega_{TP} \quad (5g)$$

- *Electrical Vehicle Taxis*

Due to growing penetration of EVTs in power systems, it is necessary to consider these modern vehicles in expansion planning studies. In this paper, EVTs penetration is included in both transmission and distribution levels. In transmission level, it is assumed that EVTs can travel between cities. With this in mind, the constraints in (6a)-(6d) are introduced. In (6a), the number of all in-service EVTs chargers is bounded according to each EVTCS capacity. The total required hourly energy for EVTs charging at each planning stage considering operational hours of EVTCSs in candidate service zones, EVTs penetration growth factor during the planning horizon, and expected EVTs charging pattern is defined by (6b). The constraints (6c) and (6d) guarantee the availability of total in-service chargers and each constructed EVTCS till the end of planning horizon.

$$\sum_{m \in \Omega_M} XE_{t,i,ze,h,m}^{TS} \leq XS_{t,i,ze}^{TS} . EP_{i,ze} \qquad \forall t \in \Omega_T, i \in \Omega_{cs}, ze \in \Omega_{ZE}, h \in \Omega_H \quad (6a)$$

$$(Ef_h . TE_h) \times (1 + Eg_t) \times (SoC^{DT} - SoC^{AT}) \leq$$
$$\sum_{ze \in \Omega_{ZE}} \sum_{i \in \Omega_{cs}} EZ_{i,ze,h}^{TS} . \left( \sum_{m \in \Omega_M} CR_m . XE_{t,i,ze,h,m}^{TS} \right) \qquad \forall t \in \Omega_T, h \in \Omega_H \quad (6b)$$

$$\sum_{h \in \Omega_H} \sum_{m \in \Omega_M} XE_{t-1,i,ze,h,m}^{TS} \leq \sum_{h \in \Omega_H} \sum_{m \in \Omega_M} XE_{t,i,ze,h,m}^{TS} \qquad \forall t \in \Omega_T, i \in \Omega_{cs}, ze \in \Omega_{ZE} \quad (6c)$$

$$XS_{t-1,i,ze}^{TS} \leq XS_{t,i,ze}^{TS} \qquad \forall t \in \Omega_T, i \in \Omega_{cs}, ze \in \Omega_{ZE} \quad (6d)$$

- *Existing and New Lines*

The constraint of (7a) defines the limits of existing line flow. In (7b), the power flow of each existing line is determined. The constraints of (7c) and (7d) model the limits and the power flow of new lines, respectively. In (7e), the availability of constructed lines at next stages is guaranteed.

$$-Pe_l^{max} \leq Pe_{t,l,h} \leq Pe_l^{max} \qquad \forall t \in \Omega_T, l \in \Omega_{el}, h \in \Omega_H \qquad (7a)$$

$$Pe_{t,l,h} = \sum_{i \in \Omega_B} \Psi \cdot B_l \cdot A_i^l \cdot \theta_{t,i,h} \qquad \forall t \in \Omega_T, l \in \Omega_{el}, h \in \Omega_H \qquad (7b)$$

$$-Pl_l^{max} \cdot Y_{t,l,c} \leq Pl_{t,l,c,h} \leq Pl_l^{max} \cdot Y_{t,l,c} \qquad \forall t \in \Omega_T, l \in \Omega_{nl}, c \in \Omega_C, h \in \Omega_H \qquad (7c)$$

$$-\mathcal{M} \cdot (1 - Y_{t,l,c}) \leq Pl_{t,l,c,h} - \sum_{i \in \Omega_B} \Psi \cdot B_l \cdot K_i^l \cdot \theta_{t,i,h} \leq \mathcal{M} \cdot (1 - Y_{t,l,c}) \qquad (7d)$$

$$\forall t \in \Omega_T, l \in \Omega_{nl}, c \in \Omega_C, h \in \Omega_H$$

$$Y_{t-1,l,c} \leq Y_{t,l,c} \qquad \forall t \in \Omega_T, l \in \Omega_{nl}, c \in \Omega_C \qquad (7e)$$

- *Power Balance Equation*

The transmission power balance equation is defined in (8). This equation includes the output power of thermal generation units and WPP considering wind curtailment, BES power exchange, power flow of existing and new lines, the exchange power between transmission and distribution systems in interface nodes, total load and EVTs charging demand.

$$P_{t,i,h} + [Wf_h \cdot Pw_{t,i} - PC_{t,i,h}] + \left[\sum_{tp \in \Omega_{TP}} (Pd_{t,i,tp,h} - Pc_{t,i,tp,h})\right] - \sum_{l \in \Omega_{el}} A_i^l \cdot Pe_{t,l,h} - \qquad (8)$$

$$\sum_{l \in \Omega_{nl}} \sum_{c \in \Omega_C} K_i^l \cdot Pl_{t,l,c,h} = \sum_{d \in \Omega_D} IN_{d,i}^{TS} \cdot PTD_{d,t,h} +$$

$$((1 + Lg_t)^{2t} \cdot Lf_h \cdot Ld_i^{PK}) + \sum_{ze \in \Omega_{ZE}} \sum_{m \in \Omega_M} CR_m \cdot XE_{t,i,ze,h,m}^{TS} \qquad \forall t \in \Omega_T, i \in \Omega_B, h \in \Omega_H$$

### B.2. Distribution Level Constraints

Generally, a linearized DistFlow approach in radial distribution systems makes it possible to calculate the active and reactive power flows considering the voltage drop form the point of common coupling to the node of feeders. Therefore, DistFlow equations are used to model distribution level constraints with a faster computation [27].

- *Thermal DGs*

To model the operation and expansion of TDGs in distribution systems, the constraints of (9a)-(9f) are presented. The active and reactive output powers of TDGs are bounded in (9a) and (9b), respectively. The linearization of nonlinear cost function of TDGs is considered in (9c) and (9d). To limit TDGs construction in each stage, the tunnel limit constraint is introduced in (9e). In (9f), the availability of constructed TDGs at next stages is guaranteed.

$$0 \leq PG_{d,t,n,h} \leq PG_n^{max} \cdot XG_{d,t,n} \qquad \forall d \in \Omega_D, t \in \Omega_T, n \in \Omega_{nt}, h \in \Omega_H \qquad (9a)$$

$$QG_n^{min} \cdot XG_{d,t,n} \leq QG_{d,t,n,h} \leq QG_n^{max} \cdot XG_{d,t,n} \qquad \forall d \in \Omega_D, t \in \Omega_T, n \in \Omega_{nt}, h \in \Omega_H \qquad (9b)$$

$$0 \leq PS_{d,t,n,h,s}^{DS} \leq XG_{d,t,n} \cdot \left(PG_n^{max}/S\right) \qquad \forall d \in \Omega_D, t \in \Omega_T, n \in \Omega_{nt}, h \in \Omega_H, s \in \Omega_S \qquad (9c)$$

$$PG_{d,t,n,h} = \sum_{s \in \Omega_S} PS_{d,t,n,h,s}^{DS} \qquad \forall d \in \Omega_D, t \in \Omega_T, n \in \Omega_{nt}, h \in \Omega_H \qquad (9d)$$

$$\sum_{n \in \Omega_{nt}} (XG_{d,t,n} - XG_{d,t-1,n}) \leq TL \qquad \forall d \in \Omega_D, t \in \Omega_T \qquad (9e)$$

$$XG_{d,t-1,n} \leq XG_{d,t,n} \qquad \forall d \in \Omega_D, t \in \Omega_T, n \in \Omega_{nt} \qquad (9f)$$

- *Wind DGs and RPS Policy*

The RPS policy is also considered for WDGs in distribution systems by (10a)-(10c). The capacity of installed WDGs is bounded in (10a). The minimum penetration of WDGs to supply the distribution system load in each stage, is defined in (10b) according to RPS policy. In (10c) the accessibility of each installed WDGs till the end of planning horizon is ensured.

$$0 \leq Pwg_{d,t,n} \leq Pwg_n^{max} \qquad \forall d \in \Omega_D, t \in \Omega_T, n \in \Omega_{nw} \qquad (10a)$$

$$[\alpha \times t/T] \times (1 + Lg_t)^{2t} \cdot \sum_{n \in \Omega_N} Ld_{d,n}^{pk} \leq \sum_{n \in \Omega_{nw}} Pwg_{d,t,n} \qquad \forall d \in \Omega_D, t \in \Omega_T \qquad (10b)$$

$$Pwg_{d,t-1,n} \leq Pwg_{d,t,n} \qquad \forall d \in \Omega_D, t \in \Omega_T, n \in \Omega_{nw} \qquad (10c)$$

- *Electrical Vehicle Taxis*

The constraints of (11a)-(11d) are introduced to model distribution level EVTs and EVTCSs. The number of all in-service EVTs chargers is bounded in (11a). In (11b), the hourly total required energy for EVTs charging at each planning stage is considered with regard to operational hours of EVTCSs in candidate service zones. In (11c) and (11d) the availability of total in-service chargers and constructed EVTCSs till the end of planning horizon are guaranteed in each distribution system.

$\sum_{m \in \Omega_M} XE^{DS}_{d,t,n,zd,h,m} \leq XS^{DS}_{d,t,n,zd} \cdot EP_{n,zd}$ $\quad \forall d \in \Omega_D, t \in \Omega_T, n \in \Omega_{ns}, zd \in \Omega_{ZD}, h \in \Omega_H$ (11a)

$(Ef_h \cdot TE_h) \times (1 + Eg_t) \times (SoC^{DT} - SoC^{AT}) \leq$

$\sum_{zd \in \Omega_{ZD}} \sum_{n \in \Omega_{ns}} EZ^{DS}_{d,n,zd,h} \cdot (\sum_{m \in \Omega_M} CR_m \cdot XE^{DS}_{d,t,n,zd,h,m})$ $\quad \forall d \in \Omega_D, t \in \Omega_T, h \in \Omega_H$ (11b)

$\sum_{h \in \Omega_H} \sum_{m \in \Omega_M} XE^{DS}_{d,t-1,n,zd,h,m} \leq \sum_{h \in \Omega_H} \sum_{m \in \Omega_M} XE^{DS}_{d,t,n,zd,h,m}$

$\quad \forall d \in \Omega_D, t \in \Omega_T, n \in \Omega_{ns}, zd \in \Omega_{ZD}$ (11c)

$XS^{DS}_{d,t-1,n,zd} \leq XS^{DS}_{d,t,n,zd}$ $\quad \forall d \in \Omega_D, t \in \Omega_T, n \in \Omega_{ns}, zd \in \Omega_{ZD}$ (11d)

- *Existing and New Feeders*

The power flow of existing and new constructed feeders, considering the possibility of operational switching for probable reconfigurations under radiality constraints, is defined in (12a)-(12g). Active and reactive power flows of existing and new feeders are bounded by the maximum apparent power flow in (12a), (12b), (12d), and (12e). The constraints in (12c) and (12f), relate the voltage drop of nodes to active and reactive power flow of existing and new feeders considering their conductance and susceptance, that are stemmed from linearized DistFlow equations [27]. In (12g), nodal voltage magnitude is bounded.

$-SF_f \cdot XO_{d,t,f,h} \leq Pe^{DS}_{d,t,f,h} \leq SF_f \cdot XO_{d,t,f,h}$ $\quad \forall d \in \Omega_D, t \in \Omega_T, f \in \Omega_{ef}, h \in \Omega_H$ (12a)

$-SF_f \cdot XO_{d,t,f,h} \leq Qe_{d,t,f,h} \leq SF_f \cdot XO_{d,t,f,h}$ $\quad \forall d \in \Omega_D, t \in \Omega_T, f \in \Omega_{ef}, h \in \Omega_H$ (12b)

$-\mathcal{M} \cdot (1 - XO_{d,t,f,h}) \leq [\sum_{n \in \Omega_N} A^f_{d,n} \cdot V_{d,t,n,h}] + 2 \times [(Pe^{DS}_{d,t,f,h} / \Psi \cdot G_f) - (Qe_{d,t,f,h} / \Psi \cdot B_f)] \leq$ (12c)

$\mathcal{M} \cdot (1 - XO_{d,t,f,h})$ $\quad \forall d \in \Omega_D, t \in \Omega_T, f \in \Omega_{ef}, h \in \Omega_H$

$-SF_f \cdot XO_{d,t,f,h} \leq Pf_{d,t,f,h} \leq SF_f \cdot XO_{d,t,f,h}$ $\quad \forall d \in \Omega_D, t \in \Omega_T, f \in \Omega_{nf}, h \in \Omega_H$ (12d)

$-SF_f \cdot XO_{d,t,f,h} \leq Qf_{d,t,f,h} \leq SF_f \cdot XO_{d,t,f,h}$ $\quad \forall d \in \Omega_D, t \in \Omega_T, f \in \Omega_{nf}, h \in \Omega_H$ (12e)

$-\mathcal{M} \cdot (1 - XO_{d,t,f,h}) \leq [\sum_{n \in \Omega_N} K^f_{d,n} \cdot V_{d,t,n,h}] + 2 \times [(Pf_{d,t,f,h} / \Psi \cdot G_f) - (Qf_{d,t,f,h} / \Psi \cdot B_f)] \leq$ (12f)

$\mathcal{M} \cdot (1 - XO_{d,t,f,h})$ $\quad \forall d \in \Omega_D, t \in \Omega_T, f \in \Omega_{nf}, h \in \Omega_H$

$$(V^{min})^2 \leq V_{d,t,n,h} \leq (V^{max})^2 \qquad \forall d \in \Omega_D, t \in \Omega_T, n \in \Omega_N, h \in \Omega_H \quad (12g)$$

- *Radiality Constraints*

In order to guarantee the radiality of each distribution system configuration considering switching possibility of existing and new feeders, the constraints of (13a)-(13e) are presented. In (13a) the open/close state of switchable existing feeders is considered. Note that for fix existing feeders the open/close state is always equal to 1. In (13b) the open/close state of new feeder is bounded by the feeder construction state. The constraint of (13c) ensures that if a new node is connected to the system, it can't be separated as an island. Based on graph theory [28], the total closed switch feeders in each hour of operation are considered to be equal to the total number of existing and new connected nodes minus one, according to (13d). The availability of new constructed feeders, and new connected nodes is guaranteed till the end of planning horizon by the constraints of (13e).

$$XO_{d,t,f,h} \leq 1 \qquad \forall d \in \Omega_D, t \in \Omega_T, f \in \Omega_{ef}, h \in \Omega_H \quad (13a)$$

$$XO_{d,t,f,h} \leq X_{d,t,f} \qquad \forall d \in \Omega_D, t \in \Omega_T, f \in \Omega_{nf}, h \in \Omega_H \quad (13b)$$

$$\sum_{f \in \Omega_{nf}} |K_{d,n}^f| \cdot XO_{d,t,f,h} = XN_{d,t,n} \qquad \forall d \in \Omega_D, t \in \Omega_T, n \in \Omega_{NN}, h \in \Omega_H \quad (13c)$$

$$\sum_{f \in \Omega_F} XO_{d,t,f,h} = N + \sum_{n \in \Omega_{NN}} XN_{d,t,n} - 1 \qquad \forall d \in \Omega_D, t \in \Omega_T, h \in \Omega_H \quad (13d)$$

$$X_{d,t-1,f} \leq X_{d,t,f} \qquad \forall d \in \Omega_D, t \in \Omega_T, f \in \Omega_{nf}$$

$$XN_{d,t-1,n} \leq XN_{d,t,n} \qquad \forall d \in \Omega_D, t \in \Omega_T, n \in \Omega_{NN} \quad (13e)$$

- *Demand Response*

The linearized DRP in distribution systems is introduced in (14a)-(14g). In (14a) and (14b), the upper and lower bounds of DRP, that indicate the positive and negative load shifting, are restricted according to the hourly load demand in each responsive load nodes. The DRP limits are bounded in (14c). In (14d), the summation of DRP in all hours is considered to be zero. The constraint of (14e) ensures that the positive and negative load shifting cannot be activated at the same time. The positive and negative load shifting between hours considering related cross-hour price elasticity and load levels, are defined in (14f) and (14g).

$$0 \leq DU_{d,t,nd,h} \leq [(1 + Lg_t)^{2t}.Lf_h .Ld_{nd}^{pk}].ZD_{d,t,nd,h}^{+} \quad \forall d \in \Omega_D, t \in \Omega_T, nd \in \Omega_{ND}, h \in \Omega_H \quad (14a)$$

$$0 \leq DL_{d,t,nd,h} \leq [(1 + Lg_t)^{2t}.Lf_h .Ld_{nd}^{pk}].ZD_{d,t,nd,h}^{-} \quad \forall d \in \Omega_D, t \in \Omega_T, nd \in \Omega_{ND}, h \in \Omega_H \quad (14b)$$

$$-DL_{d,t,nd,h} \leq D_{d,t,nd,h} \leq DU_{d,t,nd,h} \quad \forall d \in \Omega_D, t \in \Omega_T, nd \in \Omega_{ND}, h \in \Omega_H \quad (14c)$$

$$\sum_{h \in \Omega_H} D_{d,t,nd,h} = 0 \quad \forall d \in \Omega_D, t \in \Omega_T, nd \in \Omega_{ND} \quad (14d)$$

$$ZD_{d,t,nd,h}^{+} + ZD_{d,t,nd,h}^{-} \leq 1 \quad \forall d \in \Omega_D, t \in \Omega_T, nd \in \Omega_{ND}, h \in \Omega_H \quad (14e)$$

$$-\mathcal{M}.(1 - ZD_{d,t,nd,h}^{+}) \leq DU_{d,t,nd,h} - \sum_{hd \in \Omega_H} \gamma_h^{hd} \times [\frac{(DD_{d,t,nd,h} + DD_{d,t,nd,hd})}{2} \times$$

$$\frac{(\mathcal{L}_{d,t,hd} - A\mathcal{L}_{d,t})}{A\mathcal{L}_{d,t}}] \leq \mathcal{M}.(1 - ZD_{d,t,nd,h}^{+}) \quad \forall d \in \Omega_D, t \in \Omega_T, nd \in \Omega_{ND}, h \in \Omega_H \quad (14f)$$

$$-\mathcal{M}.(1 - ZD_{d,t,nd,h}^{-}) \leq DL_{d,t,nd,h} - \sum_{hd \in \Omega_H} \gamma_h^{hd} \times [\frac{(DD_{d,t,nd,h} + DD_{d,t,nd,hd})}{2} \times \frac{(\mathcal{L}_{d,t,h} - A\mathcal{L}_{d,t})}{A\mathcal{L}_{d,t}}] \leq$$

$$\mathcal{M}.(1 - ZD_{d,t,nd,h}^{-}) \quad \forall d \in \Omega_D, t \in \Omega_T, nd \in \Omega_{ND}, h \in \Omega_H \quad (14g)$$

$$DD_{d,t,nd,h} = (1 + Lg_t)^{2t}.Lf_h .Ld_{d,nd}^{pk}$$

- *Power Balance Equation*

The nodal active and reactive power balance equations in each distribution system are defined in (15) and (16). The active power balance equation in (15) includes the exchange active power between transmission and distribution in interface nodes, the active power of TDGs and WDGs, active flow of existing and new feeders, active load considering DRP, and EVTs charging demand. The reactive power of TDGs, reactive flow of existing and new feeders, and reactive load are included in (16).

$$IN_{d,n}^{DS}.PTD_{d,t,h} + PG_{d,t,n,h} + [Wf_h.Pwg_{d,t,n}] - \sum_{f \in \Omega_{ef}} A_{d,n}^{f}.Pe_{d,t,f,h}^{DS} - \sum_{f \in \Omega_{nf}} K_{d,n}^{f}.Pf_{d,t,f,h} = \quad (15)$$

$$[((1 + Lg_t)^{2t}.Lf_h .Ld_{d,n}^{pk}) + D_{d,t,n,h}] + \sum_{zd \in \Omega_{ZD}} \sum_{m \in \Omega_M} CR_m.XE_{d,t,n,zd,h,m}^{DS}$$

$$\forall d \in \Omega_D, t \in \Omega_T, n \in \Omega_N, h \in \Omega_H$$

$$QG_{d,t,n,h} - \sum_{f \in \Omega_{ef}} A_{d,n}^{f}.Qe_{d,t,f,h} - \sum_{f \in \Omega_{nf}} K_{d,n}^{f}.Qf_{d,t,f,h} = ((1 + Lg_t)^{2t}.Lf_h .QLd_{d,n}^{pk}) \quad (16)$$

$$\forall d \in \Omega_D, t \in \Omega_T, n \in \Omega_N, h \in \Omega_H$$

## C. Benders Dual Decomposition

In this subsection the MILP formulations presented in subsections A and B, are reformulated to be solved using the BDD algorithm. The problem is decomposed into a Master Problem (MP), and two Dual Sub-Problems (DSPs), one DSP for transmission level (i.e., TDSP) and one DSP for distribution level (i.e., DDSP). In MP, the integer decision variables are optimized. In TDSP and DDSP the feasibility or optimality of MP solution for transmission and distribution systems, along with optimization of the system operation, WPP and WDG investment costs, as linear continuous variables, are evaluated. In the following, an identical compact form is defined for the objective function in (1), and (1a)-(1d), as well as for the constraints of (2a)-(16).

$$\text{Min } I_{TS}^T Y^{TS} + J_{TS}^T W^{TS} + O_{TS}^T P^{TS} + I_{DS}^T Y^{DS} + J_{DS}^T W^{DS} + O_{DS}^T P^{DS} \tag{17}$$

s.t.

$$A^{TS} Y^{TS} \geq B^{TS}, \quad A^{DS} Y^{DS} \geq B^{DS} \tag{18}$$

$$C^{TS} W^{TS} + D^{TS} P^{TS} + E^{TS} Q^{TS} = F^{TS} \quad : \sigma^{TS} \tag{19}$$

$$G_1^{TS} Y^{TS} + H_1^{TS} W^{TS} + K_1^{TS} P^{TS} + L_1^{TS} Q^{TS} = M^{TS} \quad : \lambda^{TS} \tag{20}$$

$$G_2^{TS} Y^{TS} + H_2^{TS} W^{TS} + K_2^{TS} P^{TS} + L_2^{TS} Q^{TS} \geq N^{TS} \quad : \mu^{TS} \tag{21}$$

$$C^{DS} W^{DS} + D^{DS} P^{DS} + E_1^{DS} Q^{DS} = F_1^{DS} \quad : \sigma^{DS} \tag{22}$$

$$E_2^{DS} Q^{DS} = F_2^{DS} \quad : \vartheta^{DS} \tag{23}$$

$$G_1^{DS} Y^{DS} + H_1^{DS} W^{DS} + K_1^{DS} P^{DS} + L_1^{DS} Q^{DS} = M^{DS} \quad : \lambda^{DS} \tag{24}$$

$$G_2^{DS} Y^{DS} + H_2^{DS} W^{DS} + K_2^{DS} P^{DS} + L_2^{DS} Q^{DS} \geq N^{DS} \quad : \mu^{DS} \tag{25}$$

$Y^{TS}, Y^{DS} \in \{0,1\}, \quad P^{TS}, P^{DS}, W^{TS}, W^{DS} \geq 0, \quad Q^{TS}, Q^{DS}: free$

$Y^{TS} = \{Y, XS^{TS}, XE^{TS}, I, U, J\}, \quad Y^{DS} = \{X, XG, XO, XN, XS^{DS}, XE^{DS}, ZD^+, ZD^-\},$

$P^{TS} = \{P, PS^{TS}, Pd, Pc, E, R, P\mathcal{C}\}, \quad P^{DS} = \{PG, PS^{DS}, DU, DL\},$

$W^{TS} = \{Pw\}, \quad W^{DS} = \{Pwg\},$

$Q^{TS} = \{\theta, Pe^{TS}, Pl\}, \quad Q^{DS} = \{V, QG, PTD, Pe^{DS}, Qe, Pf, Qf, D\}$

$\sigma^{TS}, \sigma^{DS}, \vartheta^{DS}, \lambda^{TS}, \lambda^{DS}: free, \quad \mu^{TS}, \mu^{DS} \geq 0$

The objective function in (17) represents (1), and (1a)-(1d). The constraint of (18) stands for (5g), (6c), (6d), and (7e), along with (9e), (9f), (11c), (11d), (13a)-(13e), and (14e). The constraint of (19) denotes (8). The constraint of (20) represents (2b) and (5e). The constraint of (21) indicates the constraints of (2a), (2c), (2d), (3a)-(3e), (4a)-(4c), (5a)-(5d), (5f), (6a), (6b), and (7a)-(7d). The constraints of (22) and (23) represent (15) and (16), respectively. The constraint of (24) corresponds to (9d), and (14d). The constraints of (9a)-(9c), (10a)-(10c), (11a), (11b), (12a)-(12g), (14a)-(14c), (14f), and (14g) are compacted in (25). The compact dual variables $\sigma^{TS}, \lambda^{TS}, \mu^{TS}, \sigma^{DS}, \vartheta^{DS}, \lambda^{DS}$ and $\mu^{DS}$ are defined for the constraints (19), (20), (21), (22), (23), (24), and (25), respectively. $I_{TS}, J_{TS}, I_{DS}$ and $J_{DS}$ are the vectors for investment cost. $O_{TS}$ and $O_{DS}$ are the vectors of operation cost. The coefficients of $A, B, C, D, E, F, G_1, G_2, H_1, H_2, K_1, K_2, L_1, L_2, M$ and $N$ are relevant matrices. $T$ is the transpose operator.

- *Master Problem*

The formulations of MP are expressed as follows:

$$Min \ Z_{lower} \tag{26}$$

s.t.

$$Z_{lower} \geq I_{TS}^T Y^{TS} + I_{DS}^T Y^{DS} \tag{27}$$

$$Z_{lower} \geq I_{TS}^T Y^{TS} + \left[F^{TS^T}\overline{\sigma^{TS}} + M^{TS^T}\overline{\lambda^{TS}} + N^{TS^T}\overline{\mu^{TS}}\right]^{(v)} + \overline{\pi^{TS}}^{(v)} \cdot \left(Y^{TS} - \overline{Y^{TS}}^{(v-1)}\right) + \\ I_{DS}^T Y^{DS} + \left[F_1^{DS^T}\overline{\sigma^{DS}} + F_2^{DS^T}\overline{\vartheta^{DS}} + M^{DS^T}\overline{\lambda^{DS}} + N^{DS^T}\overline{\mu^{DS}}\right]^{(v)} + \overline{\pi^{DS}}^{(v)} \cdot \left(Y^{DS} - \overline{Y^{DS}}^{(v-1)}\right) \tag{28}$$

$$\left[F^{TS^T}\overline{\sigma^{TS}} + M^{TS^T}\overline{\lambda^{TS}} + N^{TS^T}\overline{\mu^{TS}}\right]^{(v)} + \overline{\pi^{TS}}^{(v)} \cdot \left(Y^{TS} - \overline{Y^{TS}}^{(v-1)}\right) \leq 0 \tag{29}$$

$$\left[F_1^{DS^T}\overline{\sigma^{DS}} + F_2^{DS^T}\overline{\vartheta^{DS}} + M^{DS^T}\overline{\lambda^{DS}} + N^{DS^T}\overline{\mu^{DS}}\right]^{(v)} + \overline{\pi^{DS}}^{(v)} \cdot \left(Y^{DS} - \overline{Y^{DS}}^{(v-1)}\right) \leq 0 \tag{30}$$

& (18)

The (26) is the objective function of MP that is also the problem Lower Bound (LB). The investment cost of binary decision variables is presented in (27). The optimality cut and feasibility cuts of transmission and distribution systems are introduced using (28), (29), and (30), respectively. $v$ indicates the iteration number. $\pi^{TS}$ and $\pi^{DS}$ are the dual variables of the constraints given by (31) as auxiliary constraints for the sub-problems.

$$IY^{TS}{}_{sp} = \overline{Y^{TS}} \qquad : \pi^{TS}, \qquad IY^{DS}{}_{sp} = \overline{Y^{DS}} \qquad : \pi^{DS}$$
$$I: \text{Identity Matrix}, \qquad \qquad \pi^{TS}, \pi^{DS}: free \tag{31}$$

- *Transmission Dual Sub-Problem*

In the following, the linear programming formulation of TDSP is presented.

$$Max \ F^{TS^T}\sigma^{TS} + M^{TS^T}\lambda^{TS} + N^{TS^T}\mu^{TS} + \overline{Y^{TS}}^T\pi^{TS} \tag{32}$$

s.t.

$$D^{TS^T}\sigma^{TS} + K_1^{TS^T}\lambda^{TS} + K_2^{TS^T}\mu^{TS} \leq O_{TS} \tag{33}$$

$$C^{TS^T}\sigma^{TS} + H_1^{TS^T}\lambda^{TS} + H_2^{TS^T}\mu^{TS} \leq J_{TS} \tag{34}$$

$$G_1^{TS^T}\lambda^{TS} + G_2^{TS^T}\mu^{TS} + I\pi^{TS} \leq 0 \tag{35}$$

$$E^{TS^T}\sigma^{TS} + L_1^{TS^T}\lambda^{TS} + L_2^{TS^T}\mu^{TS} = 0 \tag{36}$$

The MP solution determines the integer decision variables (i.e., $\overline{Y^{TS}}$). According to TDSP solution, if the solution is bounded, the optimality cut (28) is formed; otherwise, the feasibility cut (29) is constructed by Modified TDSP (MTDSP).

- *Distribution Dual Sub-Problem*

The DDSP is given by (37) to (41).

$$Max \ F_1^{DS^T}\sigma^{DS} + F_2^{DS^T}\vartheta^{DS} + M^{DS^T}\lambda^{DS} + N^{DS^T}\mu^{DS} + \overline{Y^{DS}}^T\pi^{DS} \tag{37}$$

s.t.

$$D^{DS^T}\sigma^{DS} + K_1^{DS^T}\lambda^{DS} + K_2^{DS^T}\mu^{DS} \leq O_{DS} \tag{38}$$

$$C^{DS^T}\sigma^{DS} + H_1^{DS^T}\lambda^{DS} + H_2^{DS^T}\mu^{DS} \leq J_{DS} \tag{39}$$

$$G_1^{DS^T}\lambda^{DS} + G_2^{DS^T}\mu^{DS} + I\pi^{DS} \leq 0 \tag{40}$$

$$E_1^{DS}\sigma^{DS} + E_2^{DS}\vartheta^{DS} + L_1^{DS^T}\lambda^{DS} + L_2^{DS^T}\mu^{DS} = 0 \tag{41}$$

After reaching a bounded solution for TDSP, the DDSP is solved. Based on DDSP solution, if the solution is bounded, the optimality cut (28) is completed, and Upper Bound (UB) of the problem is computed as follows.

$$UB = F^{TS^T}\sigma^{TS} + M^{TS^T}\lambda^{TS} + N^{TS^T}\mu^{TS} + \overline{Y^{TS}}^T\pi^{TS} + F_1^{DS^T}\sigma^{DS} + F_2^{DS^T}\vartheta^{DS} + M^{DS^T}\lambda^{DS}$$
$$+N^{DS^T}\mu^{DS} + \overline{Y^{DS}}^T\pi^{DS} + I_{TS}^T\overline{Y^{TS}} + I_{DS}^T\overline{Y^{DS}} \qquad (42)$$

Otherwise, if the solution of DDSP is unbounded, the feasibility cut (30) is generated by solving Modified DDSP (MDDSP).

- *Modified TDSP and DDSP*

To handle unbounded conditions in TDSP and DDSP, and eliminate the extreme rays, MTDSP and MDDSP are defined. The objective functions for MTDSP and MDDSP are assumed as (32) and (37). The constraints are (33)-(36) and (38)-(41) all with a right-hand-side equals to zero.

After each iteration, if the pre-set stopping criterion in (43) is satisfied, the algorithm will be terminated. Note that in the proposed BDD algorithm $\tau$ is assumed as 0.01.

$$\frac{(UB-LB)}{UB} \leq \tau \qquad (43)$$

### III. OUTLINE OF THE PROPOSED MODEL

The overall structure of proposed coordinated model contains three main parts. First part is MP in which the integer decision variables are optimized. Other two main parts are TDSP and DDSP. In TDSP and DDSP the feasibility or optimality of MP solution for transmission and distribution systems, along with optimization of the system operation, WPP and WDG investment costs, as linear continuous variables, are evaluated. The outline of proposed coordinated model is displayed in Fig. 3. As it can be seen, to start the proposed BDD algorithm, firstly, the input data and initial values are provided. Then, MP (i.e., (26)-(30), and (18)) is solved and binary decision variables are obtained and considered as fixed inputs to TDSP and DDSP. After solving TDSP (i.e., (32)-(36)), if the solution is unbounded, MTDSP is solved, and the feasibility cut (29) is generated for MP. Then, the next iteration is started. However, if the solution of TDSP is bounded, the optimality cut (28) is generated for MP, and LMPs in interface buses are sent to DDSP. It should be noted that LMPs are calculated using dual variable of (8), i.e., $\sigma^{TS}$. Then, DDSP (i.e., (37)-(41)) is solved. According to DDSP solution, if the solution is unbounded, MDDSP is solved and the feasibility cut (30) is generated for next iteration MP. On the contrary, if the solution of DDSP is bounded, the optimality cut (28) is completed. After sending the exchange power (i.e., $PTD$) to TDSP, the UB (42) will be calculated. The algorithm is terminated if the criterion in (43) is satisfied; otherwise, the next iteration is started.

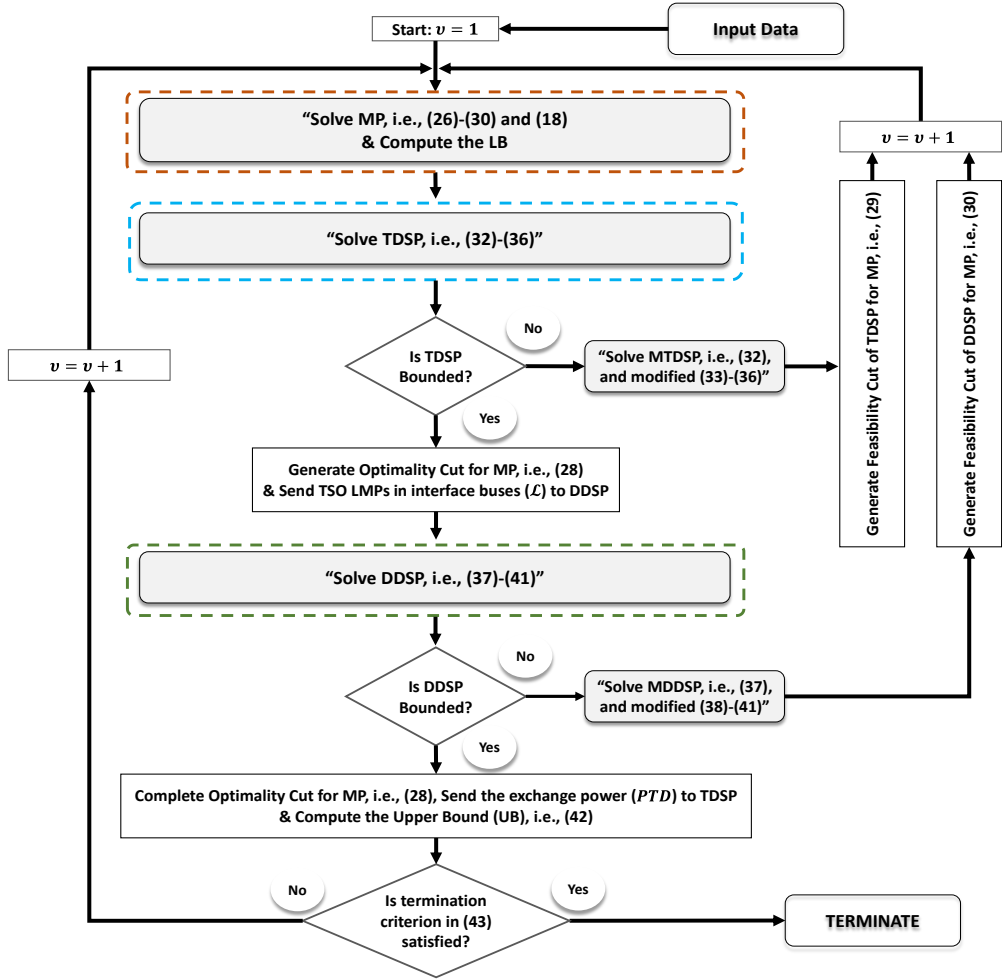

Fig. 3. Outline of the proposed model

## IV. REPRESENTATIVE PERIODS

To decrease the computational complexity of system operation modeling, and capture the short-term uncertainty of load demand, WPPs and WDGs output power, and the charging pattern of EVTs in each station, an accurate CTPC algorithm [29], is utilized. The prerequisite for detailed modeling of operational details in a long-term planning is to preserve the temporal chronology with a sufficiently high resolution. The CTPC algorithm can keep the chronology (inter-temporal feature) of time-varying parameters and consider the correlation between load and EVTs charging demand, and wind power. Therefore, under high penetration of WPPs and WDGs, the short-term uncertainties can be captured more accurately in a long-term horizon. The historical data of load demand [30], wind power [31], and the charging pattern of EVTs [32], in Netherlands at 2019 are considered to extract the representative hours. Moreover, the aggregated historical data are accessible in [33]. Due to lack of EVTs charging pattern data in whole hours of a year, MCS method is utilized to extend the available data over a year considering the data probability distribution. The normal probability distribution function is considered for EVTs charging pattern available data. As shown in Fig. 4, by using the CTPC algorithm, the hourly load demand, EVTs charging pattern, and wind power historical data are represented by 96 hours. The extracted representative hours are aggregated across the year using each representative weight.

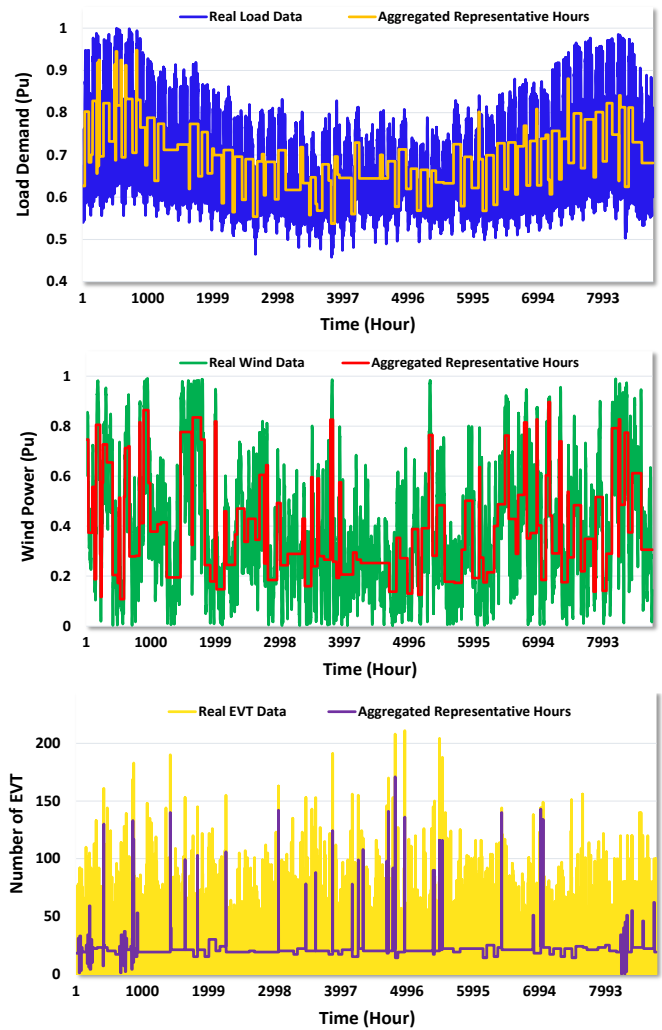

Fig. 4. The real historical data and extracted representative hours

V. NUMERICAL RESULTS

The performance of the proposed coordinated expansion planning model of transmission and active distribution systems is evaluated over the IEEE 24-bus test system, as transmission system, along with four IEEE 33-node test feeders, as distribution systems. Four distribution systems are connected to buses 22, 17, 12, and 11 of considered transmission system. The data of these test systems at the beginning of the planning horizon are assumed as input data and extracted from MATPOWER [34]. The planning time horizon is divided into 3 stages; each stage contains 2 years. In this paper, the generation expansion planning is not considered. Indeed, it is assumed that enough generation capacity is available in the system until the end of considered planning horizon. The investment costs of $IC$, $Rw$, and $TSb$ are 1 M$/Km, 0.034 M$/Km, and 3.358 M$, respectively [29]. The costs of $Cs$ and $Cc$ are considered as 0.5 M$/MW, and 0.05 M$/MWh, and the $\eta_c$ and $\eta_d$ are considered as 0.9 [29]. $IW$ and $Cwc$ are assumed as 2 M$/MW, and 2000 $/MWh [29]. The $ES$, $EC$, $CR$, and $TE$ are 1000 $, 60 $, 0.05 MW, and 0.0028 MWh, respectively [35]. The percentages of $\alpha$, $\beta$, and $\chi$ are equal to 20%, 35%, and 10%, respectively. In distribution systems, $IF$, $DR$, $DSb$, $DT$, and $DW$ are

considered equal to 0.02 M$/Km, 0.007 M$/Km, 0.95 M$, 0.98 M$, and 0.65 M$/MW, respectively [36]. The $Lg$ in transmission and distribution systems, and $r$ are equal to 5%. The $Eg$ for the first, second, and third stages is assumed as 2%, 11%, and 30% [35]. All input data and parameters are also accessible in [33]. The simulations are run by the CPLEX solver in GAMS software using a PC with an Intel Core i7 CPU, and 32 GB of RAM.

In IEEE 24-bus test system, six WPPs are installed in buses 3, 6, 12, 14, and new buses 25, and 26. The maximum capacity of candidate WPPs in existing buses is assumed as 180 MW, and 300 MW for new buses. The buses 6, 10, 14, 16, 18, 20 and 21 are considered as candidates for three types of BES devices. In this paper, three different types of BES devices are considered as candidates. The maximum power capacity of candidate BES devices is 25, 50, and 100 MW, with a maximum energy capacity of 150, 300, and 600 MWh. Four zones are considered for EVTCSs candidate service zones in transmission level. The zones include buses 6, 8, 10 (zone 1), 9, 11, 12 (zone 2), 20, 22, 23 (zone 3), 3, 15, and 24 (zone 4). In all feeders of IEEE 33-node test system, four TDG with maximum capacity of 3 MW are considered to be constructed in nodes d3, d5, d11, and d29. The nodes d16, d22, d30, d34, and d35 are candidates for WDG installation. The maximum capacity of WDGs is 0.5, 1, 1, 2.5, and 2.5 MW, respectively. The candidate service zones in distribution systems are nodes d5, d6 (zone 1), d16, d17 (zone 2), d31, and d32 (zone 3). The responsive load nodes are d2, d4, d7, d8, d14, d18, d19, d24, d26, and d32. The numerical results are given in two parts to confirm the efficacy of proposed model as follows.

*A. Numerical Results for Uncoordinated Planning*

In this subsection, the numerical results of the proposed model excluding the coordination between transmission and distribution systems are presented. In this regard,

Table II. Results of The Uncoordinated Planning

| Stage | New Line (From-To) | WPP (Max. Cap. MW) | BES (MW/MWh) | EVTCS (Zone) |
|---|---|---|---|---|
| 1 | 16-19, 7-8 | **3**: (105), **6**: (180) | **6**: (100/600), **10**: (25/150), (50/300), **14**: (25/150), (50/300), **16**: (50/300), (100/600), **18**: (25/150), (100/600), **21**: (25/150), (50/300), (100/600) | **6 & 8**: (1), **9 & 11**: (2), **22**: (3), **15 & 24**: (4) |
| 2 | 21-25 | **3**: (75), **14**: (180), **25**: (102) | — | — |
| 3 | 9-11, 10-12, 21-25, 2×(16-26) | **25**: (164.85), **26**: (255.55) | **10**: (100/600) | **12** (2) |
| $IC^{1,2}$: | 465.24 | 534.23 | 292.6 | 0.051 |
| **Total Costs:** | $TTIC^3$: 1292.12 | | $TTOC^4$: 3079.4 | $Z^5$: 4371.52 |

1: Investment cost, 2: All costs are in **M$**, 3: Total transmission investment cost,
4: Total transmission operation cost, 5: Total planning cost distribution systems are assumed as a simple bus with/without load demand and generation in interface buses. As reported in Table II, in the first stage, two lines between buses 7-8, and 16-19 are constructed. In this stage, two WPPs in buses 3 and 6, also twelve BES devices are installed in buses 6, 10, 14, 16, 18, and 21, as presented in Table II. Moreover, seven EVTCSs are constructed in buses 6, 8, 9, 11, 22, 15, and 24 in first stage. In second stage, one line in corridor 21-25 is constructed. Additionally, three WPPs are installed in buses 3, 14, and 25. In the last stage of planning, three lines are constructed between buses 9-11, 10-12, 21-25, and two lines between buses 16-25. Two WPPs in buses 25, and 26, along with one BES device in bus10, are installed in third stage. Moreover, one EVTCS is located in bus 12. As presented in Table II, TTIC, TTOC, and total planning cost are obtained as 1292.12, 3079.4, and 4371.52 $M\$$, respectively. A total wind energy of 56.05 GWh is curtailed until the end of planning horizon in the proposed model excluding the coordination between transmission and distribution systems.

### *B. Numerical Results for Coordinated Planning*

The numerical results of the proposed model considering the coordination between transmission and four distribution systems connected to buses 11, 12, 17, and 22 are

Table III. Results of The Proposed Coordinated Planning Excluding DRP

| TS[1] | New Line (From-To) | WPP (Max. Cap. MW) | BES (MW/MWh) | EVTCS (Zone) |
|---|---|---|---|---|
| ST[2] 1 | 16-19, 7-8 | **3**: (106), **6**: (180) | **6**: (100/600), **10**: (25/150), (50/300), **14**: (25/150), (50/300), **16**: (50/300), (100/600), **18**: (50/300), (100/600), **21**: (50/300), (100/600) | **8**: (1), **9**: (2), **22**: (3), **24**: (4) |
| ST 2 | 21-25 | **3**: (74), **14**: (180), **25**: (94) | — | — |
| ST 3 | 9-11, 10-12, 21-25, 2×(16-26) | **25**: (172), **26**: (242.5) | **10**: (100/600) | **11**: (2) |
| *IC*: | 465.24 | 527.63 | 292.6 | 0.0301 |
| **Total Costs:** | ***TTIC***: 1285.5 | | ***TTOC***: 3006.5 | ***TTPC***[3]: 4292 |

| System | Investment Cost (M$) | | | |
|---|---|---|---|---|
| | New Feeder | TDG | WDG | EVTCS |
| **DS1**[4] | 0.551 | 1.313 | 2.033 | 0.0024 |
| **DS2** | 0.551 | 1.313 | 2.033 | 0.0024 |
| **DS3** | 0.551 | 1.313 | 2.033 | 0.0024 |
| **DS4** | 0.551 | 1.313 | 2.033 | 0.0024 |
| **Total Costs:** | ***TDIC***: 15.597 | ***TDOC***: 29.153 | ***TDPC***[5]: 44.75 | **Z**: 4336.75 |

**1:** Transmission system, **2**: Stage, **3:** Total transmission planning cost,
**4**: Distribution system, **5:** Total distribution systems planning cost

presented in this subsection. All the connecting buses contain 12 MW load demand. To illustrate the efficiency of the proposed model, the simulations are conducted for two cases. In **case I**, the proposed coordinated model excluding DRP is executed, while in **case II** DRP is also incorporated in the proposed model. The numerical results of cases I, and II are reported in Tables III, and IV, respectively. Moreover, the obtained results of case II are illustrated in Fig. 5. As given in Table III, the values of decision variables are just presented for transmission system. In addition, all investment and operation costs are distinguished for both transmission and distribution systems. In case I, the TTIC, TTOC, and total transmission planning cost (TTPC) are 1285.5, 3006.5, and 4292 *M$*, respectively. In comparison to the results in Table II, which is an uncoordinated model, TTPC is reduced by 79.52 *M$*. This reduction in TTPC confirms the importance of the proposed coordinated planning model. As shown in Table III, the TDIC, TDOC, and total distribution planning cost (TDPC) are 15.597, 29.153, and 44.75 *M$*, respectively. The

total planning cost in case I, as the summation of TTPC and TDPC, is 4336.75 *M$* that leads to a 34.77 *M$* cost saving compared to the obtained result of Table II. The total wind energy curtailment in case I during the planning horizon is 44.18 GWh which is 11.87 GWh less than the curtailed energy in uncoordinated planning model. The obtained cost saving confirms the effectiveness of proposed coordinated planning model. Note that the cost saving is obtained just by the interaction of four distribution systems with the transmission grid. By modelling more distribution systems, the cost saving regarding the

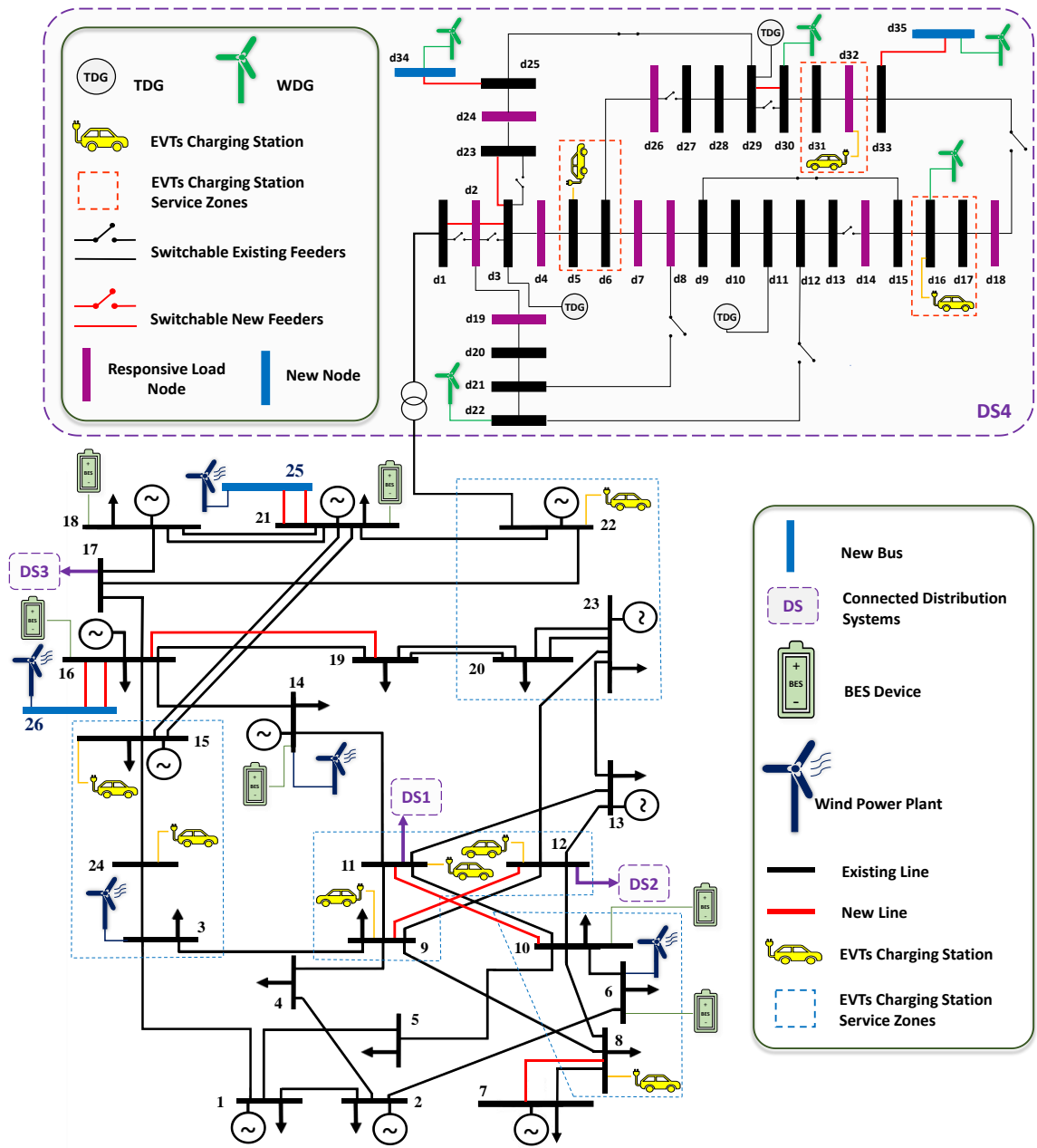

Fig. 5. The illustration of the obtained results of case II

uncoordinated planning model can be increased.

According to Table IV, by considering DRP in case II, the TTIC, TTOC, and TTPC are 1292.5, 2998.65, and 4291.16 $M\$$, respectively. Moreover, in case II the TDIC, TDOC, and TDPC are 14.89, 24.765, and 39.65 $M\$$, respectively. The total planning cost (i.e., $Z$) in case II (i.e., Table IV) is 4330.81 $M\$$ that includes 4291.16 $M\$$ as TTPC, and 39.65 $M\$$ as TDPC. The total planning cost of case II is 5.94, and 40.71 $M\$$ less expensive than case I (i.e., Table III) and the results of uncoordinated model (i.e., Table II), respectively. TTPC in case II is 80.36 $M\$$ less expensive than the results of Table II. The obtained results of case II confirm the influence of DRP on reducing both distribution and transmission total planning and operation costs. As illustrated in Fig. 5, in case II eight new lines, five WPPs, six BES devices, and seven EVTCSs are constructed in transmission system. In addition, six new feeders, three TDGs, five WDGs, and three EVTCSs are constructed in the fourth distribution system, i.e., the system connected to bus 22. In Fig. 6, the total energy balance for all distribution systems in case II, including TDGs and WDGs output energy, energy supply from transmission system, energy demand, and EVT charging demand, is illustrated in each planning stage. Based on Fig. 6, in stages one, two, and three, the share of TSO in suppling DSO energy consumption is 59%, 55%, and 43%, respectively. This result confirms the importance of proposed coordination between transmission and distribution systems.

Table IV. Results of The Proposed Coordinated Planning Including DRP

| TS | New Line | WPP | BES | EVTCS |
|---|---|---|---|---|
| $IC$: | 465.24 | 527.63 | 299.59 | 0.049 |
| **Total Costs:** | $TTIC$: 1292.5 | $TTOC$: 2998.65 | $TTPC$: **4291.16** | |

| System | Investment Cost (M$) | | | |
|---|---|---|---|---|
| | New Feeder | TDG | WDG | EVTCS |
| DS1 | 0.552 | 1.193 | 2.033 | 0.0024 |
| DS2 | 0.552 | 1.193 | 2.033 | 0.0024 |
| DS3 | 0.552 | 1.084 | 2.028 | 0.0024 |
| DS4 | 0.552 | 1.084 | 2.030 | 0.0024 |
| **Total Costs:** | $TDIC$: 14.89 | $TDOC$: 24.765 | $TDPC$: **39.65** | $Z$: **4330.81** |

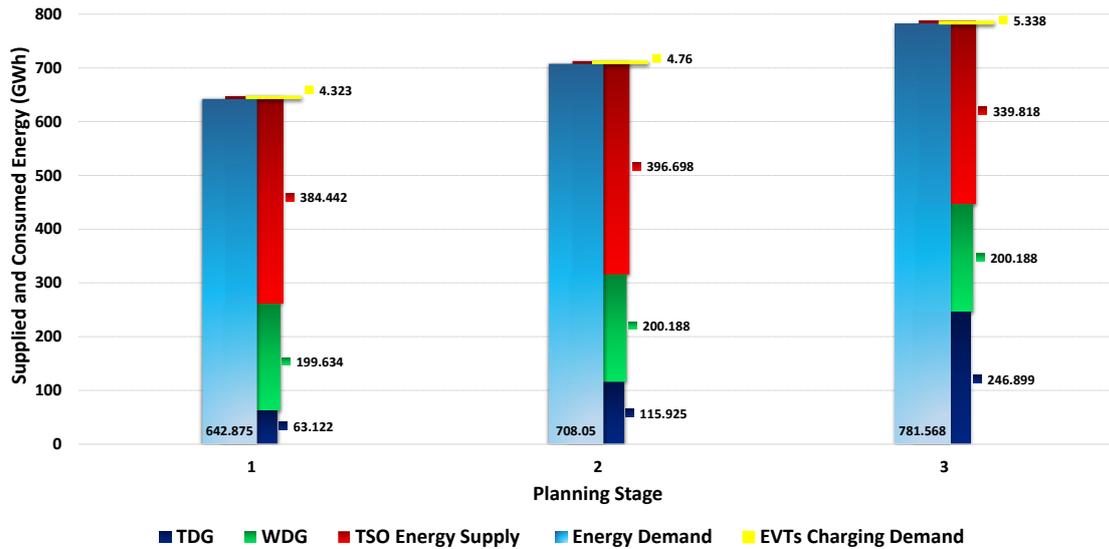

Fig. 6. The total energy balance for distribution systems in case II

It should be noted that in order to evaluate the effect of forecasting error in the proposed model, it is possible to simulate the considered case studies for different probable scenarios of forecasted parameters. In this paper, some sensitivity analyses can be conducted over load demand and EVT growth factors, the penetration of WPPs in the last stage of the planning horizon, and interest rate. Although the focus of this paper is on modelling the integration of smart grid technologies in active distribution systems planning, and their direct impacts on transmission network planning, flexible ramp reserve of thermal units is incorporated in the proposed model. Indeed, flexible ramp reserve is considered to cover the probable forecast errors and handle the uncertainty of load demand and wind power plants output power as much as possible.

## VI. CONCLUSION

This paper proposed a stochastic multistage model to coordinate expansion planning of transmission and active distribution systems, concerning short-term operational details. In the transmission level, lines, BES devices, along with WPPs, and in the distribution level, switchable feeders, TDGs, and WDGs were considered as planning options. The expansion of EVTCSs was modeled in both levels, and the impact of DRPs, implemented by DSO, was planned and scheduled in the proposed model. The proposed coordinated model results in a reduction in total expansion planning cost that means social welfare is improved. Such a cost saving is obtained just by the interaction of four distribution systems with the transmission grid; obviously it would be even more for a large number of distribution systems. The obtained results under modeling DRP, implemented by DSO with regard to TSO LMPs in interface buses, confirm the DRP influence on the reduction of total planning and operation costs. The examined operational hours of EVTCSs in candidate service zones lead to near realistic results for modeling the EVTs charging pattern. Without considering the operational details, the impacts of smart grid technologies on transmission and distribution systems are not truly captured. The proposed BDD algorithm is able to deal with the different complexities of the proposed coordinated planning model with numerous decision variables. As mentioned, the effectiveness of proposed coordinated model will be more impressing with considering the interaction of more distribution systems with transmission system in next studies. In addition, the proposed model in this paper can be discussed in future works considering an electricity market environment. Moreover, in future investigations, the utilized CTPC method in this paper can be improved to capture both temporal chronology and extreme values of input data inside some proper representatives.